\documentstyle[11pt,newpasp,twoside,epsf]{article}
\def\edcomment#1{\iffalse\marginpar{\raggedright\sl#1\/}\else\relax\fi}
\reversemarginpar

\begin{document}

\title{System to estimate ages and redshifts for radio galaxies}

\author{
    O.V.Verkhodanov,
    A.I.Kopylov,
    N.V.Verkhodanova,
    O.P.Zhelenkova,
    V.N.Chernenkov,
    Yu.N.Parijskij,
    N.S.Soboleva,
    A.V.Temirova
}
\affil{Special astrophysical observatory, Nizhnij Arkhyz, Russia}

\begin{abstract}
The system allowing a user
to operate at server with simulated curves of spectral
energy distributions (SED) and
to estimate ages and redshifts by photometric data {\bf sed.sao.ru}
is described.
\end{abstract}

Appearing of huge volume of observational data
in the optical and infrared wavelength range increases significantly
our knowledge about the far Universe.
However, information about space distribution of extragalactic
objects is not yet accessible because of the limits of observational
possibilities: the direct measuring of redshifts is possible
with spectroscopic methods having a sensitivity 2 magnitude lower
than photometric ones.
The special interest lays in the study of distant objects allowing
astronomers
to investigate both Universe structures and
evolution of active galactic nuclei (AGNs) (which are connecting with
black holes).
Spectroscopy of
such
objects
is rather difficult.
However, using photometric data one can essentially simplify
this problem since it allows an astronomer to make the initial selection.

%
To accelerate a procedure of age (and photometric redshift)
estimation we have begun a project ``Evolution of radio galaxies'',
which has to allow a user to obtain
age and photometric redshift estimations.
The main tasks of the system are:
1)    estimation of ages when fixed redshift $z$;
2)    estimation of both ages and $z$;
3)    archiving of optical observations of RC radio galaxies
     (in FITS, JPEG, PS formats with text comments);
5)    archiving of the main publications by the current topic;
6)    developing of the HTPP and e-mail access;
7)    local SED operation to simulate an observational process.

To estimate ages and redshifts by photometry data we
operate with simulated curves of spectral energy distributions (SED)
for different types of galaxies
of two models
    PEGASE (Fioc, Rocca-Volmerange, 1997, 1999),
    GISSEL96 (Bruzual, Charlot, 1993; Bolzonella et al., 2000).

Before the estimation of values we smooth a SED with a filter transmission
curve to simulate observational data using a ``compressing'' filter with
the growth of redshift:
$
S_{ik} = \frac{\sum_{j=0}^n s_{i-n/2+j} f_{jk}(z)} {\sum_{j=0} f_{jk}(z)} .
$
Here
$s_i$ is the initial synthetic SED,
$S_{ik}$ is the smoothed SED by the $k$-th filter,
$f_k(z)$ is the transmission in the k-th filter, ``compressed''
	    in $(1+z)$ times when ``moving'' along the SED,
$j =1,n$ is the pixel index in a curve of filter transmission.

The estimation of ages and redshifts is performed by way of selection of the
optimum location on the SED curves
of the measured photometric points obtained when observing
radio galaxies in different filters. We use the already
computed table SED curves for different ages.
Using discrepancies we construct a probability function in the form
of
$
	p = \frac{1}{max}exp(\chi^2),
$
where
$max$ is the maximum value of the calculated function.
$\chi$ is the discrepancy calculated by the slipping
of the photometry points along the SED curve:
$$
  \chi^2 = \sum_{k=1}^{N filters} \left( \frac{F_{obs,k} - p SED_{k}(z)}{\sigma_k}\right)
$$
Here
$F_{obs,k}$ is the observational magnitude in the $k$--th filter,
${\tt SED}_{k}(z)$ is the simulated magnitude for the given SED
in the $k$--the filter at the given redshift $z$,
$p$ is the free coefficient, $\sigma_k$ is the error of
the observed magnitude.

  In order to take account of the absorption, we apply the maps (as
FITS-files
from
the paper ``Maps of Dust IR Emission for Use in Estimation of Reddening and
CMBR Foregrounds'' (Schlegel et al., 1998).

    The system is situated on the special Web-server {\bf http://sed.sao.ru}
operating in the Linux Red Hat (6.2) system,
unifying various type resources and accessed
by FTP, HTTP and e-mail.
Typical e-mail form of request looks like this:
\begin{verbatim}
seds start
object 3C65; model PEGASE,  type=E
z_limits: 0 6,    age_limits: 200 16000
B=23.73+0.21 V=23.57+0.2 R=22.36 I=20.81
extinction off
seds end
\end{verbatim}
Here {\tt seds start} and {\tt seds end} are opening and closing
keywods of the form.  Keywords {\tt object}, {\tt model} and {\tt type}
determine an object name, a type of a model and a type of a galaxy,
respectively. Available galaxy types are E, S0, Sa, Sb, Sc, Sd.
{\tt z\_limits} and {\tt age\_limits} determine the limits of search
for a redshift and an age (in Myr).
The observed magnitudes age given with B, V, R, I, J, H, K, g, r, i, etc.
keywods corresponding to the filter names.
The error of the magnitude detection is given via plus '+' after
a value of the magnitude.
Extinction in this example is not calculated.

Another supported possibilities are
  a)sorted bibliographical collection of papers for different stages of
       radio galaxy evolution;
  b)  archive of radio galaxies data in various
       wavelength ranges (both observed in Special astrophysical observatory
       and taken from Internet), containing information on the
       objects and figures in FITS, JPEG and PostScript formats.

The project was supported by the Russian Foundation of Basic Researches
(grant No 99-07-90334).


\end{document}